\documentclass[12pt]{iopart}
\usepackage{iopams}
\usepackage[square,sort&compress]{natbib} 
\usepackage{graphicx} 
\begin{document}
\title{Application of the Gillespie algorithm to a granular intruder particle}
\author{J. Talbot$^1$ and P. Viot$^2$}
\address{$^1$Department of  Chemistry  and  Biochemistry,
 Duquesne University, Pittsburgh, PA 15282-1530}
\address{$^2$Laboratoire de Physique
Th\'eorique des Liquides, Universit\'e Pierre et Marie Curie, 4, place
Jussieu, 75252 Paris Cedex, 05 France}

\begin{abstract}
We show how the  Gillespie algorithm, originally developed to describe
coupled  chemical     reactions, can be    used  to  perform numerical
simulations  of  a   granular  intruder particle    colliding with
thermalized  bath particles.   The algorithm  generates  a sequence of
collision ``events'' separated  by variable time intervals.  As input,
it requires  the position-dependent  flux of  bath particles  at  each
point on the surface of the intruder particle.  We validate the method
by applying  it   to a one-dimensional  system   for  which the  exact
solution of   the  homogeneous  Boltzmann    equation  is  known   and
investigate the  case where the bath particle velocity distribution  has
algebraic tails.  We also present  an application to a granular needle
in bath of  point  particles where  we   demonstrate the presence   of
correlations between the    translational and rotational degrees    of
freedom of the    intruder  particle.  The relationship    between the
Gillespie algorithm  and the  commonly  used  Direct Simulation  Monte
Carlo (DSMC) method is also discussed.

\end{abstract} 

\submitto{\JPA}
\section{Introduction}

Kinetic theories of  granular systems are usually constructed starting
from the  Boltzmann  equation or one of  its variants\cite{GG01,PB03}.
Rarely, it  is possible to  obtain an exact,  analytic solution of the
Boltzmann      equation\cite{PTV06}.    More   typically,     however,
approximations are required. It is then highly desirable to assess the
quality of the theoretical  prediction  by comparing it with  accurate
numerical  solutions of the Boltzmann  equation. It  is the purpose of
this  paper  to show  that, besides the  celebrated  Direct Simulation
Monte  Carlo (DSMC)  introduced  by  Bird\cite{B94}, there  exists  an
alternative method, originally  proposed by Gillespie\cite{G76,G77} to
study coupled chemical reactions.

One class of system that is amenable to a  Boltzmann approach and that
has  received  considerable attention in  recent   years consists of a
single   intruder (or   tracer) particle   in  a bath  of  thermalized
particles\cite{DB05,BRM05,MP99,VT04,GTV05,PVTW06},    showing       in
particular the   absence of  equipartition.  This  phenomena  has been
observed  experimentally    in    two-dimensional\cite{FM02}       and
three-dimensional\cite{WP02}      granular gases.  The   intruder-bath
particle collisions are dissipative,  while the bath particles have an
ideal  gas    structure  and    a   specified   velocity  distribution
characterized  by  a   fixed   temperature. Since  Gaussian   velocity
statistics  are  quite  rare,   it is  important  that  numerical  and
theoretical approaches be able to treat a general distribution.

We begin by outlining the Gillespie  algorithm and how  it can be used
to obtain  a  numerical solution of the   Boltzmann equation.  We then
illustrate the application  of the algorithm with  two  examples.  The
first is a one-dimensional  system consisting of an intruder  particle
in a bath of thermalized point particles.  If  the bath particles have
a  Gaussian velocity  distribution  one  recovers  the exact  Gaussian
intruder  velocity  distribution function \cite{MP99}  with a granular
temperature   that is smaller than  the  bath temperature. In addition
when   we  impose  a  power-law   velocity   distribution on the  bath
particles, we find   the    same  form for the    intruder    particle
distribution.

The second system is two-dimensional and consists of a needle intruder
in a bath of point particles.  It is known that equipartition does not
hold between different degrees of freedom of the needle\cite{VT04}.
Here we use the Gillespie algorithm to obtain a new physical result,
namely the presence of correlations between the translational and
rotational degree of freedom of the needle.

\section{Algorithm}
The  model consists of  a  single intruder  particle that  undergoes a
series of collisions with  the surrounding bath particles. Let  $P(t)$
denote the probability  that no event  (collision) has  occurred in the
interval $(0,t)$. Then
\begin{equation}
P(t+\Delta t)=P(t)(1-\phi(t)\Delta t+O(\Delta t^2))
\end{equation}
where $\phi(t)$ is the event rate in general and the collision rate in
this application. Expanding the lhs to first order in $\Delta t$ and taking
the limit $\Delta t\to0$ leads to
\begin{equation}
\frac{d\ln P(t)}{dt}=-\phi(t)
\end{equation}
Integrating and using the boundary condition that
$P(0)=1$ gives
\begin{equation}\label{eq:wait}
P(t)=\exp\bigl[-\int_0^{t} \phi(t')dt'\bigr]
\end{equation}
If $\phi(t)$ is constant between collisions this 
expression takes the simple form:
\begin{equation}\label{eq:waits}
P(t)=\exp(-\phi t)
\end{equation}
At the end of the waiting time an event (collision) occurs that alters
the value  of $\phi$ or   the way  that $\phi$  evolves   with time. We  now
consider  two   specific  applications. In the   first,    the flux of
colliding particles  is  constant between  collisions  and the simpler
form, Eq    \ref{eq:waits}  may be used.   Our  second  example, an
anisotropic  object that rotates  between  collisions, requires the
use of the more general form, Eq \ref{eq:wait}.

\section{Applications}

\subsection{One-dimensional system}
Consider  first   a one-dimensional  system   consisting  of an  intruder
particle  of mass $M$ moving in  a bath of thermalized point particles
each  of mass $m$. The dynamics of the intruder is described by the 
Boltzmann equation. 

 The velocity  distribution of the bath particles is
denoted  by $f(v,a)$,  where $a$ is  related  to the bath temperature,
$T_B$. 

The flux of particles that  collide with  the right  hand side of  the
intruder particle moving with a velocity $v_1$ is:
\begin{eqnarray}
\phi_+(v_1)&=&\rho\int_{-\infty}^{v_1}(v_1-v)f(v,a)dv,
\end{eqnarray}
where $\rho$ is the number density of the bath particles. Similarly the flux on
the left hand side is:
\begin{eqnarray}
\phi_-(v_1)&=&\rho\int_{v_1}^{\infty}(v-v_1)f(v,a)dv
\end{eqnarray}
and the total collision rate is 
\begin{eqnarray}
\phi(v_1)&=&\phi_+(v_1)+\phi_-(v_1)
\end{eqnarray}

Let $F(v_1,t)$ denote the time-dependent distribution function of tracer 
particle velocity. Then this evolves according to
\begin{equation}\label{eq:master}
\frac{\partial F(v_1,t)}{\partial t}=-\phi(v_1)F(v_1,t)+\int_{-\infty}^{\infty}\psi(v\to v_1)F(v,t)dv
\end{equation}
which is equivalent to the homogeneous Boltzmann equation\cite{PTV06}. 
The first term on the rhs is a loss term corresponding to the
probability that a tracer particle with velocity $v_1$ undergoes a
collision (necessarily to a different velocity) per unit time. The second, or
gain, term contains the function  $\psi(v\to v_1)$ that is the rate that tracer particles moving with velocity
$v$ are transformed (by collisions with the bath particles) 
to those moving with velocity $v_1$. For collisions with the rhs of the intruder particle the explicit
expression is
\begin{equation}
\psi_+(v\to v_1)=\rho\left(\frac{1+M}{1+\alpha}\right)^2(v-v_1)f(v+\frac{1+M}{1+\alpha}(v_1-v),a),
\end{equation}
with $v_1<v$. A similar expression applies for collisions on the left hand side of the 
intruder particle. We note that a representation of the 
Boltzmann equation similar to Eq \ref{eq:master} was employed by Puglisi et al. \cite{PVTW06}. 

The Gillespie algorithm provides an numerical solution of Eq
\ref{eq:master}, {\it including} the transient case when the derivative
is not equal to zero.

If the intruder  particle moves with a constant  velocity the flux  is
itself (on   average)  constant. A waiting   time  consistent  with Eq
\ref{eq:wait} is then generated:
\begin{equation}\label{eq:deltat}
\Delta t=-\ln(\xi_1)/\phi(v_1),
\end{equation}
where $0< \xi_1<1$ is a uniform random number.

Given a collision  at time  $t$, the  probability that the   collision
occurs on the right is given by
\begin{equation}\label{eq:side}
p(+|t)=\phi_+(v_1)/ \phi(v_1).
\end{equation}
This    is sampled by  generating    a  second  uniform  random number
$0<\xi_2<1$. If $\xi_2<p(+|t)$  the collision is on  the  right hand side:
otherwise it takes place on the left hand side.

Having chosen the side, it is then necessary to sample the velocity of
the bath particle that collides with this side. The probability
distribution function of the colliding particle's velocity depends on
the collision side:
\begin{eqnarray}\label{eq:vdist1} 
g_{+}(v,v_1)=\left\{ \begin{array}{ll}
(v_1-v)f(v,a)/\phi_+(v_1) & \mbox{if $v\leq v_1$}  \\
0 & \mbox{otherwise}
\end{array}
\right. \end{eqnarray}

\begin{eqnarray}\label{eq:vdist2} 
g_{-}(v,v_1)=\left\{ \begin{array}{ll}
(v-v_1)f(v,a)/\phi_{-}(v_1) & \mbox{if $v\geq v_1$}  \\
0 & \mbox{otherwise}
\end{array}
\right. \end{eqnarray}

\begin{figure}
\begin{center}

\resizebox{6cm}{!}{\includegraphics{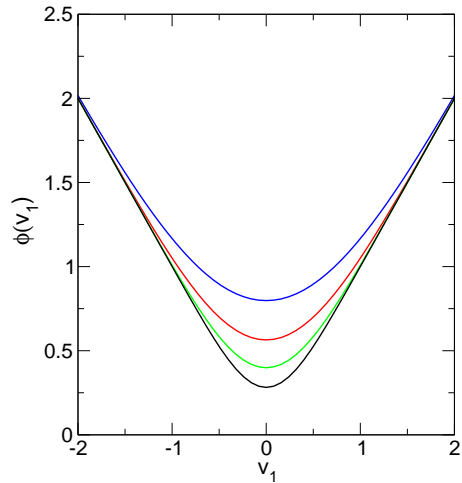}}
\caption{Flux, or collision rate, on both sides of a surface moving in a bath of particles 
with a Gaussian velocity distribution at a velocity $v_1$. $a=0.5,1,2,4$ top-to-bottom. 
}\label{fig:flux}
\end{center}
\end{figure}

The results presented so far apply to any bath velocity
distribution - and the behavior depends strongly on the 
exact form \cite{PVTW06}. For a Gaussian
\begin{equation}\label{eq:maxwell}
f(v,a)=\sqrt{\frac{a}{\pi}}\exp(-av^2),\;\;-\infty<v<\infty 
\end{equation}
where $a=m/(2k_BT_B)$. The collision fluxes on each side of the tracer particle are
\begin{equation}
\phi_{\pm}(v_1)= \frac{\rho}{2}(\pm v_1+\frac{1}{\sqrt{\pi a}}\exp(-av_1^2)+v_1{\rm erf}(v_1\sqrt{a})),
\end{equation}
and the total collision rate is 
\begin{equation}
\phi(v_1)=\rho(\frac{1}{\sqrt{\pi a}}\exp(-av_1^2)+v_1{\rm erf}(v_1\sqrt{a})).
\end{equation}
This function is shown in Figure \ref{fig:flux}.

\begin{figure}
\begin{center}

\resizebox{6cm}{!}{\includegraphics{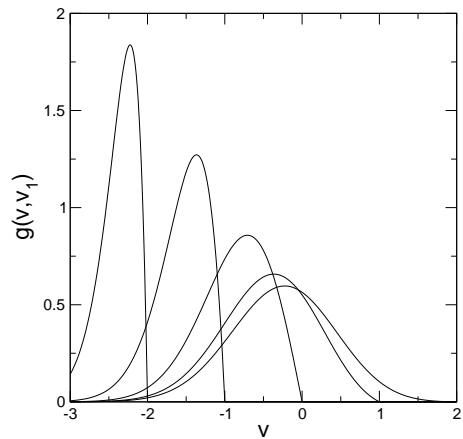}}
\caption{Probability distribution of particles that collide with the right hand side of a surface 
moving  with   a  velocity    $v_1=-2,   -1,  0,   1,   2$,  left   to
right. The bath particles have a Gaussian velocity distribution}\label{fig:vdist}
\end{center}
\end{figure}

The  velocity distribution of the colliding particles, Eq \ref{eq:vdist1}, 
in the case of a bath particle Gaussian velocity distribution  
is  plotted  in    Fig~\ref{fig:vdist}  for several values of the intruder
particle velocity.   Sampling of this
distribution is accomplished with an acceptance-rejection method 
in which the sampling region is adapted to the velocity of the intruder. 
A rare, but
possible, case is to  select a collision on  the right hand  side when
the surface   is moving rapidly  to  the  left. The   distribution  of
colliding particles is sharply peaked at $v_1$  and it is necessary to
take the range of $v$ from $v_1-2$ to $v_1$.  If the surface is moving
to the right,   the distribution of  colliding particle's  velocity is
more symmetric and one can sample $v$ in the range $-3\leq v \leq 3$.

Finally, when    the intruder particle moving   with  a velocity $v_1$
collides with  a  bath particle  of velocity  $v$ the  velocity of the
former changes instantaneously to
\begin{equation}\label{collisionrule}
v_1'=v_1+\frac{1+\alpha}{1+M}(v-v_1),
\end{equation}
where $0<\alpha\leq 1$ is the coefficient of restitution and we have taken
$m=1$ for convenience.

\begin{figure}
\begin{center}

\resizebox{6cm}{!}{\includegraphics{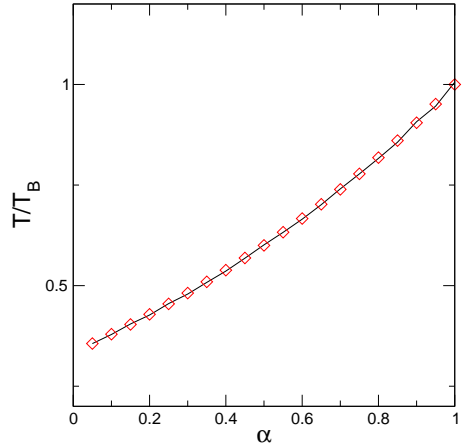}}
\caption{Ratio of the intruder granular temperature to the bath temperature. 
Symbols show the simulation results and the solid curve
shows     the theoretical prediction  of     Martin  and Piasecki,  Eq
\ref{eq:MP}.    Results      are   for      $M=m$     and      $n_{\rm
col}=500000$}\label{fig:1dresults}
\end{center}
\end{figure}

The complete algorithm describing one iteration can now be summarized using pseudocode:
\begin{itemize}

\item[1] 
Generate a waiting time using Eq \ref{eq:deltat}. 

\item[2]
$t \to t + \Delta t$\\
 while $ ( t_{\rm out} < t)$\\
$\{ t_{\rm out}  \gets  t_{\rm out}+\delta t$\\
accumulate averages$\}$

\item[3]
Choose the collision side using Eq \ref{eq:side}. 

\item[4]
The velocity of the colliding bath particle $v$  is sampled from the
distribution given by Eq \ref{eq:vdist1} or Eq \ref{eq:vdist2}. 

\item[5]
The post-collisional velocity of the intruder particle is determined from
Eq \ref{collisionrule}.

\end{itemize} 

Although the time increment between events is variable, averages (mean
square velocities and velocity distributions) must be computed at equal
time intervals. In step 2, $t$ denotes the total elapsed time and $\delta
t$ is the constant time interval between the accumulation of
quantities to be averaged. A convenient choice for $\delta t$ is the
average collision time.

Martin and Piasecki  \cite{MP99} obtained an analytic solution of the 
homogeneous
stationary  Boltzmann equation  and
showed that  the velocity distribution  of the intruder  particle in the
steady state is Gaussian and  characterized  by a temperature,  $T$,
that is different from the bath  temperature $T_B$.  Specifically, the
two are related by:
\begin{equation} \label{eq:MP}
\frac{T}{T_B}=\frac{1+\alpha}{2+(1-\alpha)/M},
\end{equation}
so that  for   $\alpha<1$, $T<T_B$.   Figure \ref{fig:1dresults} shows an 
excellent agreement  of the simulation with this exact result.

Since the existence of solutions of the Boltzmann equation with
power-law tails has been shown recently \cite{BMM05,BM05}, it is
interesting to investigate this phenomenon in the intruder particle
system using the Gillespie algorithm.  Therefore, we consider the case
where the bath particle velocity distribution function takes the
following power-law form:
\begin{equation}\label{eq:2}
f(a,v)=\frac{\sqrt{2a}}{\pi}\frac{1}{1+a^2v^4}
\end{equation}
The granular temperature of the bath is well defined since the average
of the square velocity is finite,  $<v^2>=1/a$.

Although the Boltzmann equation can no longer be solved in general for
an arbitrary bath particle velocity distribution (unlike the just-discussed
Gaussian distribution) an exact solution is possible when 
the mass ratio is equal to the coefficient of restitution, $M/m=\alpha$. For
this specific case
one can  show that the
stationary  solution  of the intruder    particle is exactly  given  by
Eq.(\ref{eq:2})\cite{PTV06} with a granular
temperature  equal to  the   bath  temperature multiplied  by the
coefficient of restitution.

\begin{figure}
\begin{center}

\resizebox{9cm}{!}{\includegraphics{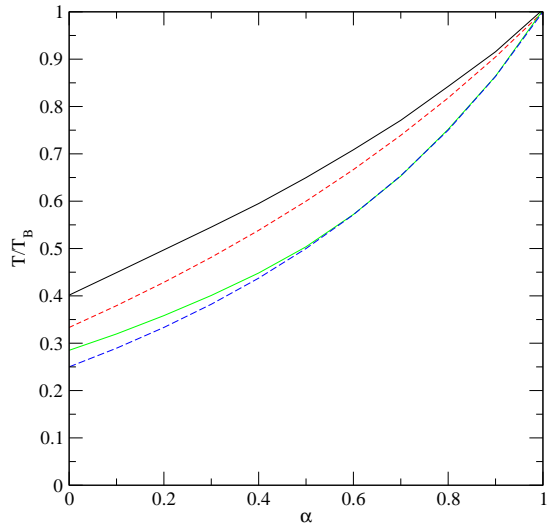}}
\caption{Granular temperatures of the intruder particle  
  normalized by the bath temperature, $T/T_B$, versus the coefficient
  of restitution $\alpha$ when the velocity distribution of the bath is
  given by Eq. (\ref{eq:2}) for $M=1$ and $M=0.5$, from top to bottom.
  Dashed curves correspond to the analytical result \cite{MP99}
  when the velocity distribution of the bath is
  Gaussian.}\label{fig:8}
\end{center}
\end{figure}

Figure~\ref{fig:8}  shows the variation of
the  granular temperature of the  intruder particle with $\alpha$ for $M/m=1$
and $M/m=0.5$.  When  the intruder is light, $M<m$, the
granular temperature  of the intruder  particle is close to the result
obtained  with     the      Gaussian  bath when     $\alpha>M/m$     (see
Fig. \ref{fig:8}).  

\begin{figure}
\begin{center}

\resizebox{9cm}{!}{\includegraphics{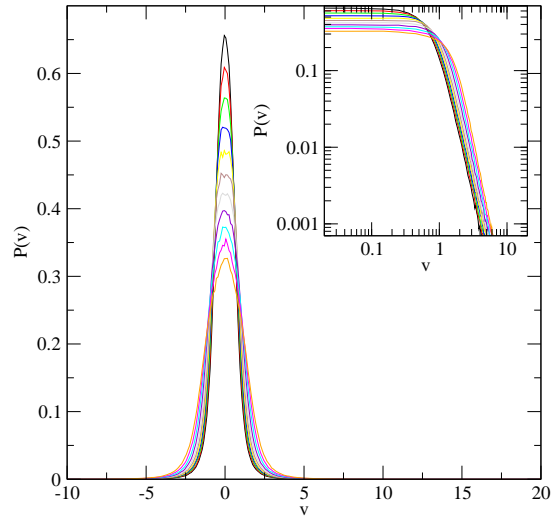}}
\caption{Velocity distribution function of the intruder particle for different values of the coefficient of restitution 
  $\alpha=0.0,0.1,0.2,...0.8,0.9,1.0$,  from top to bottom, when $M=1/2$ and when the bath
  distribution is given by Eq.(\ref{eq:2}). The insert is a log-log plot showing the algebraic decay of the velocity distribution} \label{fig:9}
\end{center}
\end{figure}

Figure \ref{fig:9} displays the intruder particle velocity
distribution function for different values of the coefficient of restitution 
$0.0\leq\alpha\leq1.0$ when $M=1/2$. The exact solution is known for $\alpha=0.5$, i.e. 
Eq.(\ref{eq:2}) with a granular temperature equal to $\alpha$.  The simulation
results show that in all cases the velocity distribution functions exhibit a
power-law tail (See inset of Fig.\ref{fig:9}) with an exponent
independent of $\alpha$ and equal to $-4$.

\subsection{Needle}

\begin{figure}[t]
\begin{center}

\resizebox{8cm}{!}{\includegraphics{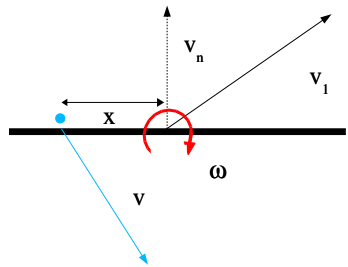}}
\caption{Geometry of the needle-point system}\label{fig:needle}
\end{center}
\end{figure}

In this application a needle intruder, confined to a
two-dimensional plane, is immersed in a fluid of point particles, each
of mass $m$, at a density $\rho$.  The needle is characterized by its
mass $M$, length $L$ and moment of inertia $I$ and its state is
specified by its angular and center of mass velocities, $\omega$ and ${\bf
  v_1}$, respectively (see Fig.\ref{fig:needle}).  The velocity distribution of the point particles
is again given by Eq \ref{eq:maxwell}. 

Two main modifications of the Gillespie algorithm are required in
order to simulate this system. First, if $v_1\neq0$ the flux is not 
(on average) constant
between collisions. Second, it is necessary to select the point of
impact on the needle.

The collision flux on both sides of the
needle at a point $-L/2\leq x\leq L/2$ is $\phi({\bf v_1}.{\bf n}+\omega x)$. Unlike
the case considered above, this flux is 
time-dependent if $\omega\neq 0$ since the normal vector ${\bf n} $ 
rotates.
Specifically
\begin{equation}\label{eq:vn}
v_n={\bf v_1}.{\bf n}(t)=v_{1y}\cos(\omega t+\theta_0)-v_{1x}\sin(\omega t+\theta_0),
\end{equation}
where $\theta_0$ is the orientation of the needle at $t=0$. 
The total flux over the entire length of the needle is
\begin{equation}\label{eq:fluxn}
\Phi(t) =\int_{-L/2}^{L/2}\phi({\bf v_1}.{\bf n}(t)+\omega x)dx.
\end{equation}
For the sake of simplicity we take $a=1$ and $L=1$ and we obtain 
\begin{eqnarray}
\Phi(t)&=&\rho \left(\frac{v_n^2}{2\omega}+\frac{\omega}{8}+\frac{v_n}{2}
+\frac{1}{4\omega}\right)erf \left(v_n+\frac{\omega}{2}\right)\nonumber\\
&+&\rho \left(\frac{v_n^2}{2\omega}+\frac{\omega}{8}-\frac{v_n}{2}
+\frac{1}{4\omega}\right)erf\left(-v_n+\frac{\omega}{2}\right)\nonumber\\
&+&\rho \frac{e^{-v_n^2-\frac{\omega^2}{4}}}{4\sqrt{\pi}}\left[e^{v_n\omega}\left(1-\frac{2v_n}{\omega}\right)+
e^{-v_n\omega}\left(1+\frac{2v_n}{\omega}\right)\right].
\end{eqnarray}

A collision time is generated by solving the equation
\begin{equation}
\int_0^tdt'\Phi(t')=-\ln(\xi_3)
\end{equation}
Since  the  integral cannot be performed    analytically, we use the
Newton-Raphson method: 
\begin{equation}\label{eq:1}
t_1=t-\frac{\int_0^t\Phi(t')dt'+\ln(\xi_3)}{\Phi(t)}
\end{equation}
with an   initial guess $t=-\frac{\ln(\xi_3)}{\Phi(0)}$.  The procedure  is
iterative,   i.e. $t$  is    substituted by $t_1$ in  Eq.(\ref{eq:1}),
etc. until $|\int_0^{t_1}\Phi(t')dt'+\ln(\xi_3)|$ is smaller than the required
precision.  In  general   the convergence  is   fast and only   a few
iterations are required.   Occasionally,   when  $\Phi$ is    small,   the
Newton-Raphson method oscillates  between two ``stable'' positions and
there is no convergence.   When this situation  arises, we switch to a
bisection  procedure. With this modification, the   method seems to be
robust.

Once the time to collision  has been selected,  one updates the normal
velocity of the center-of-mass of the needle using Eq \ref{eq:vn}.

In order to  choose the position of  the impact, one has to  calculate
the   probability that the collision occurs at a distance $x$ from 
the center of mass of the needle, whatever the velocities
of the bath particles, at a given time $t$
\begin{equation}
p(x|t)=  \frac{\Phi (v_n+x\omega)}{\Phi(v_n,\omega)},\;\;-1/2\leq x \leq 1/2
\end{equation}
This probability is clearly not uniform over the length of the needle. But
since it is concave, the maximum is obtained for $x=1/2$ or
$x=-1/2$ (if $v_n=0$ the probability is a maximum at both ends).  We
select $x$ using a standard acceptance-rejection method.

Once the position of the  impact has been decided, one  has  to choose if the
bath  particle collides on the right or left hand side of the needle. 
The probability of the former event is
\begin{equation}
p(+|x,t)=\frac{\Phi_+(v_n+x\omega)}{\Phi(v_n+x\omega)}
\end{equation}
One
chooses a random  number  between $0$ and   $1$. If $\xi<p(+|x,t)$,  the
particle collides  with  a particle  from   the right,  otherwise  the
collision is on the left.

Next the velocity of the colliding bath particle must be selected.
The probability that the colliding particle has a velocity between $v$
and $v+dv$ is $g_{\pm}(v,v_n+x\omega)dv$. It is even more important to
sample this distribution carefully than in the one-dimensional case as
more extreme velocities are encountered in this system.

Finally, the needle velocity and angular velocity are updated 
using
\begin{eqnarray}
{\bf v_1}'&=&{\bf v_1}+{\bf n}\frac{\Delta  p}{M}\\
I\omega'&=&I\omega+x\Delta p
\end{eqnarray}
where
\begin{equation}
\Delta p=-\frac{(1+\alpha){\bf g}.{\bf n}}{\frac{1}{m}+\frac{1}{M}+\frac{x^2}{I}},
\end{equation}
${\bf n}$ is a unit vector normal to the length of the needle and
${\bf g}={\bf v_1}-{\bf v}+\omega x{\bf n}$ is the relative velocity at
the point of impact (the velocity of the colliding bath particle also
changes, but we do not need to know the new value).

It is necessary to simulate a few hundred thousand to several million
collisions in order to obtain good estimates for the properties of
interest. The convergence of the simulation is slower when the mass
of the needle is much bigger than the bath particle mass.

\begin{figure}
\begin{center}

\resizebox{8cm}{!}{\includegraphics{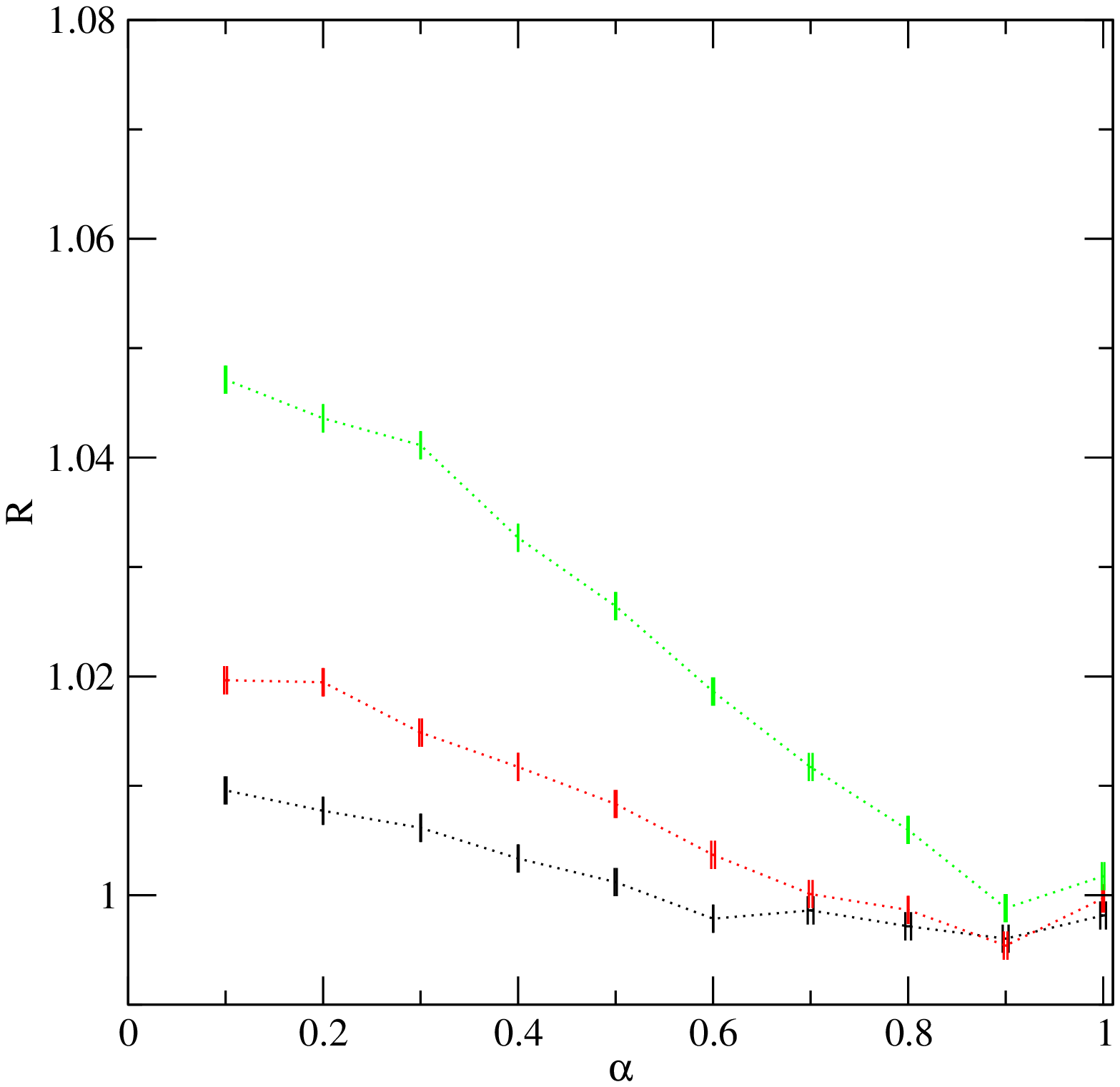}}
\caption{Correlation coefficient, Eq \ref{eq:cor}, as a function of the coefficient of restitution for $M=0.1,0.5,1$, top-to-bottom}\label{fig:R}
\end{center}
\end{figure}

We have previously used this algorithm to confirm a kinetic theory
prediction that, when the coefficient of restitution is smaller than
unity, the temperature of the bath is larger than the 
translational granular temperature which is in turn
larger than the rotational granular temperature \cite{VT04}.

Here we present new results for the 
cross-correlation between the two degrees of freedom of the needle as
a function of the coefficient of restitution, $\alpha$, and for different
values of the mass ratio $M/m$.  Specifically, we have calculated
\begin{equation}\label{eq:cor}
R=\frac{<v^2\omega^2>}{<v^2><\omega^2>},
\end{equation}
which is equal to one in equilibrium systems and obviously independent of the 
mass ratio. Conversely, one observes that for small mass ratios in an inelastic system,
there is a positive correlation that increases with decreasing $\alpha$:
See Figure \ref{fig:R}. We expect this to be a generic feature 
of anisotropic granular particles in
any dimension, regardless of the bath particle velocity distribution.

\begin{figure}
\begin{center}

\resizebox{8cm}{!}{\includegraphics{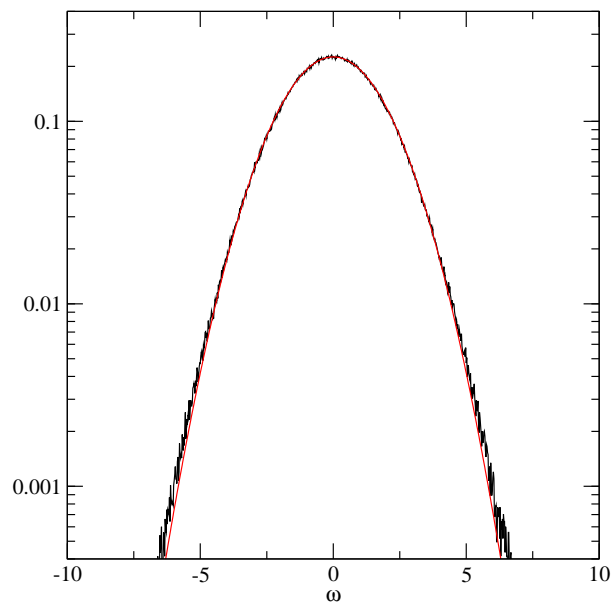}}
\caption{Angular velocity distribution for $M/m=0.1$ and $\alpha=0.1$. Black line: 
simulation with 5000000 collisions; Red line: best-fit Gaussian}\label{fig:om}
\end{center}
\end{figure}

Unlike the one-dimensional  model, where the  velocity distribution of
the intruder particle is Gaussian for all values of the coefficient of
restitution, the translational  and angular  velocity distributions of
the granular needle  are never strictly  Gaussian (except for  elastic
collisions). For a large range  of values of  $M/m$ and $\alpha$,  however,
the Gaussian is  a very accurate approximation.   It is only  for a light
needle and highly  inelastic  collisions that deviations  start to
become apparent: See Figure \ref{fig:om}.

\section{DMSC versus Gillepsie algorithm}
The two simulation methods provide an exact numerical solution of
a Boltzmann equation (they can also be used for inelastic Maxwell
models\cite{BK03,BM05,BMM05}). The DSMC algorithm has been described in 
several articles \cite{B94,MS00}. For convenience we describe here 
the version that would be
applied to the one-dimensional system discussed in section 3.1. 

Velocities sampled from a Gaussian distribution, Eq \ref{eq:maxwell},
are assigned to $n_{\rm bath}$ bath particles. A side of the intruder
particle is then selected at random: $\sigma_{i1}=+1$ for a collision on
the left, $\sigma_{i1}=-1$ for a collision on the right. A bath particle,
$i$, is then selected randomly. The collision is accepted with
probability $\Theta(v_{i1}\sigma_{i1})\omega_{i1}/ \omega_{max}$ where $\Theta(x)$
is the Heaviside function, $\omega_{i1}=2\rho |v_{i1}|$ and $\omega_{max}$ is
an upper bound estimate of the probability that a particle collides
per unit time. If the collision is accepted a post-collisional
velocity, computed from Eq \ref{collisionrule}, is assigned to the
intruder particle. If $\omega_{i1}>\omega_{max}$ the estimate of the latter is
updated: $\omega_{max}\gets\omega_{i1}$.

We have implemented this algorithm for the one-dimensional intruder
described in section 3.1. We obtain the same results with comparable
computational effort.

\section{Conclusion}
We have shown how the Gillespie algorithm can be used to 
obtain an exact numerical solution of the Boltzmann equation of
an intruder particle in a bath of particles with an arbitrary 
velocity distribution. We used the method to obtain new results
for a one-dimensional system consisting of an 
intruder particle in a bath with a power
law distribution. We also used it to demonstrate, for the first time, the presence 
of correlations between the translational and rotational momenta
of an anisotropic particle. 

Although the results presented here apply 
to the steady state, the method is equally valid for the transient
case. It is clear that the Gillespie algorithm offers no significant
computational advantage over the DSMC method for these intruder
particle systems. It is of interest, however, that two apparently
dissimilar methods can be applied to the same physical system. 
Finally, we note that, as with DSMC, the Gillespie method can be
easily generalized to three-dimensional systems.

\ack We thank Alexis Burdeau, Jaros\l aw Piasecki and Thorsten P\"oschel for helpful discussions. 
\section*{REFERENCES}
\providecommand{\newblock}{}

\end{document}